\documentclass[
    ,final            
  ,numberedheadings 
	,twocolumn
  ]
  {aipproc}
  
\usepackage{graphicx}
\usepackage{epstopdf}
\usepackage{amsmath}
\usepackage{amsfonts}
\usepackage{amssymb}

\layoutstyle{8x11double}

\begin{document}

\title{Refractive Index Matched Scanning and Detection of Soft Particles}

\classification{81.05.Rm}
\keywords      {index matching, image analysis, soft spheres, hydrogel, granular packings, aspherical}

\author{Joshua A. Dijksman}{
  address={Physical Chemistry and Soft Matter, Wageningen University, Wageningen, The Netherlands}
}

\author{Nicolas Brodu}{
  address={GeoStat, INRIA Bordeaux Sud-Ouest, Bordeaux, France}
}

\author{R.P. Behringer}{
  address={Dept. of Physics, Duke University, Science Drive, Durham NC 27708-0305, USA}
}

\begin{abstract}
We describe here how to apply the three dimensional imaging technique of refrecative index matched scanning to hydrogel spheres. Hydrogels are water based materials with a low refractive index, which allows for index matching with water-based solvent mixtures. We discuss here various experimental techniques required to handle specifically hydrogel spheres as opposed to other transparent materials. The deformability of hydrogel spheres makes their identification in three dimensional images non-trivial. We will also discuss numerical techniques that can be used in general to detect contacting, non-spherical particles in a three dimensional image. The experimental and numerical techniques presented here give experimental access to the stress tensor of a packing of deformed particles.
\end{abstract}

\maketitle


\section{Introduction}
Granular materials such as sand, rice, powders and coffee beans have surprising mechanical behavior. They are defined as collections of particles whose thermal energy is negligible, yet they can behave as liquids, solids and gases~\cite{1996_revmodphys_jaeger, 2008_annurevfluid_forterre}. Both packing fraction and shear stress have emerged as crucial variables necessary to describe the state of the granulate. One can cross the phase boundary between solid and liquid granular materials thus \emph{jamming} the granulate, either by shear~\cite{bi11} or compaction~\cite{2009_jphyscondmat_hecke}. Most of these effects have been observed in numerical simulations or quasi two dimensional experiments. However, experimental studies that explore the microstructure of granular materials in three dimensional systems are rare~\cite{brujic, saadatfar, hurley16}. Refractive index matched scanning (RIMS) has emerged as a useful tool to perform three dimensional imaging on dense granular samples. The technique has been perfected over the past decade to the extent it can now be applied in many different experimental contexts~\cite{DijksmanRSI12, budwig, wiederseiner}. In this review article, we explore how \textit{soft} transparent particles can be used for RIMS-based granular experiments. We focus in particular on hydrogel particles. Hydrogel particles can be index matched with watery solutions and are therefore particularly amenable to use with RIMS. Hydrogels, and soft particles in general, however present the experimentalist with technical challenges. For example: finding the right kind of solvent to index match watery porous gel particles requires careful considerations. The first part of this article is entirely devoted to such experimental considerations; it is an update of previous work~\cite{dijksman13}.\\
Another feature of experiments on soft particles is the need for finding the complete surface structure of particles in the packing. As the particles are soft, they deform under mechanical loading. The implication is that the usual assumption of particle sphericity breaks down. This has consequences for finding their position, as the center of mass now depends on the shape. However, measuring particle level deformations has significant benefits: obtaining shape information allows one to extract contact forces and contact networks~\cite{brodu}. In general, it allows one to extract particle level stress tensor information from packings of soft spheres. In addition the shape information of particles gives access to particle level anisotropy metrics, which are also relevant to understand the microscopic mechanics of granular packings~\cite{gerd}. The second part of this article is therefore devoted to a general discussion on how to extract particle shape information from three dimensional (3D) images. The algorithmic aspects described here are generic; they may also apply to 3D images obtained by means of confocal~\cite{brujic}, X-Ray~\cite{saadatfar} or magnetic resonance imaging methods.\\

\section{Using hydrogel spheres}

\begin{figure}
  \includegraphics[width=1\columnwidth,clip]{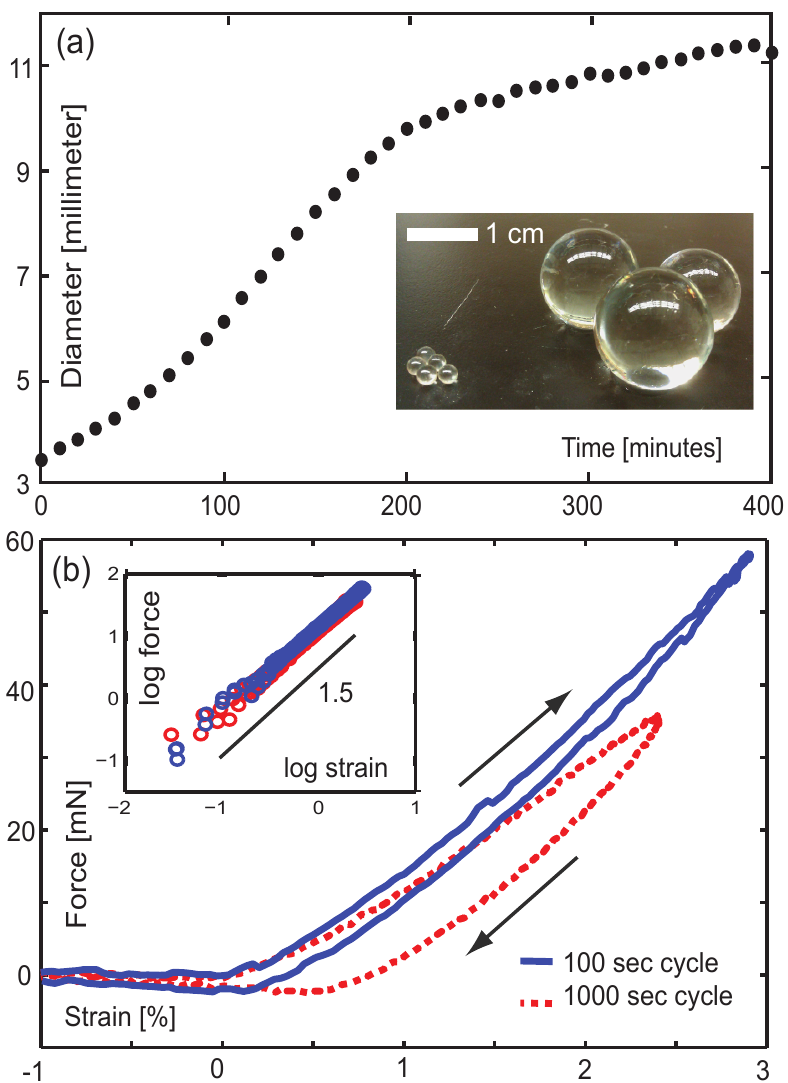}
  \caption{Reproduced from AIP Conf. Proc. \textbf{1542}, 457 (2013) with the permission of AIP Publishing. (a) Typical growth dynamics of a single dried hydrogel sphere in demineralized water. The inset shows a dry and fully grown sphere. (b) Compression force as a function of strain for a single fully grown hydrogel sphere. Total cycle time is 100 seconds (blue) and 1000 seconds (dashed red). The inset shows the same data on double 10-logarithmic scale, with a power law of exponent 1.5 indicated. \label{fig:compresssphere}}
\end{figure}

Refractive index matching studies rely on combining particles and fluids with the same refractive index. Due to the simplicity of hard sphere suspensions and availability of materials, RIMS studies are mostly done with plastic or glass particles~\cite{DijksmanRSI12, budwig, wiederseiner}, index matched with viscous and/or expensive liquids. However, spherical particles made out of hydrogel (or polyacrylamide gel) have become commercially available~\cite{footnote1} and methods to custom produce them have been proposed~\cite{workamp}. Hydrogel as a material has the benefit that it is cheap, safe and easily index matched with a watery solution. In addition, hydrogel is soft, with a Young's modulus in the 10-100 kPa range, so it allows for the detection of contacts through its macroscopic deformation under modest forcing. The softness of the particles means that care should be taken to generalize findings in such a system to systems with hard, frictional grains. However, the easy deformability does gives an opportunity to probe the role of particle shape on granular packing mechanics. In this first part of this review, we describe experimental aspects of using hydrogel spheres in RIMS studies that make imaging of hydrogel particles practically possible.  

\subsection{Swelling Dynamics} 
Commercially available hydrogel spheres~\cite{footnote1} grow as indicated in Fig.~\ref{fig:compresssphere} when submersed in demineralized water. They are purchased as millimeter sized hard spheres; their final size depends on the properties of their chemical environment such as salinity, pH and temperature~\cite{muniz, shin, verneuil}. The swelling is reversible; upon drying, the particles revert to their dry spherical shape, althrough there is an asymmetry in the swelling/drying dynamics~\cite{thibault}. We describe the properties of these spheres throughout the paper unless otherwise noted. Note that hydrogel particles can be custom made with a droplet pinch-off technique~\cite{technote, patel, workamp} or by injection molding~\cite{bellani}.
 
\subsection{Mechanical Behavior} 
\begin{figure}
  \includegraphics[width=1\columnwidth]{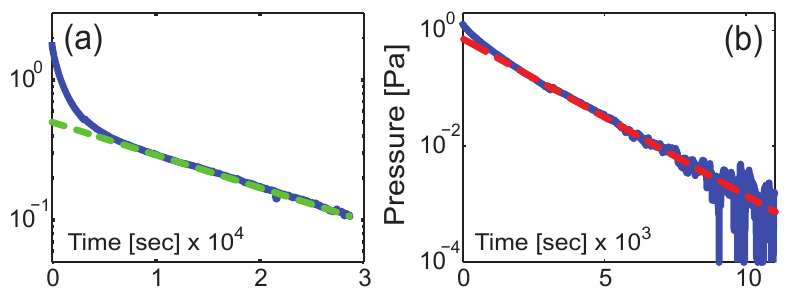}
  \caption{Reproduced from AIP Conf. Proc. \textbf{1542}, 457 (2013) with the permission of AIP Publishing. (a) Pressure evolution of a packing of hydrogel spheres kept at constant volume (blue solid line). An exponential decay to zero at late times can be observed; the green dashed line is a best fit with time constant of 5.1 hours. (b) Subtracting the late time exponential decay, an early time exponential relaxation process with a different time constant is observed (blue solid line). The red dashed line is a fit with decay time of 0.5 hour. Note the different time scales in panel (a) and (b).\label{fig:relax}}
\end{figure}

To measure the response of a single sphere, we perform compression tests in a TA Instruments Micro Strain Analyzer III.  We compress a single hydrogel sphere vertically and uniaxially between two steel plates and measure the resultant force $F$ as a function of strain $\gamma$. We use a fully swollen hydrogel sphere with no liquids on its surface to minimize liquid bridge induced capillary forces. We perform experiments in an open container and hence evaporation is present. Typical experimental results are shown in Fig.~\ref{fig:compresssphere}. The force response $F(\gamma)$ is consistent with an exponent of $1.5$, in accordance with linear elasticity theory for compressed spheres. This Hertzian behavior is also observed for submersed spheres; see below. The prefactor $Y$ in $F = Y\gamma^{\alpha}$ is related to the Young's modulus of the hydrogel. Experimental results shown in Fig.~\ref{fig:compresssphere}b suggest that the Young's modulus for the hydrogel spheres grown in demineralized water is $\sim 20~kPa$. This is in line with literature values~\cite{muniz}, and also depends on the chemical environment (e.g. salinity or pH) of the spheres as they are grown~\cite{muniz, shin}; we always used demineralized water as growing environment.\\
The hydrogel spheres are made out of cross-linked polymers which form an elastic network. At any finite temperature however, the elastic network bonds can rearrange and the interstitial fluid can move around, allowing the gel to lower its internal stresses. The hydrogel network will therefore adjust to applied stresses~\cite{stressrelax1, stressrelax2, stressrelax3}. This is also visible in Fig.~\ref{fig:compresssphere} from the finite surface area of the force loop; during the compression cycle of the sphere, the porous polymer network evolves, leading to a different force response. The pressure adjustment is more noticeable when the compression loop is slowed. To characterize this relaxation more quantitatively, we compress a packing of $\sim$1000 spheres in a fixed volume and probe the pressure response over time. The packing is fully submersed to prevent evaporation of water in the hydrogel particles. Fig.~\ref{fig:relax}a shows the typical pressure response. At late times, an exponential relaxation can be observed, with a decay timescale of about 5.1 hours. Subtracting the late time behavior reveals an exponential early time relaxation process with a considerably faster decay timescale of about 0.5 hours. This double relaxation process is robust: it is also present as a background process when the packing is sheared quasi-statically. However, when the compression level is varied stepwise as in Ref.~\cite{brodu}, the relaxation disappears. We speculate that particle level contact evolution also can play a role.\\

\begin{figure}
  \includegraphics[width=1\columnwidth]{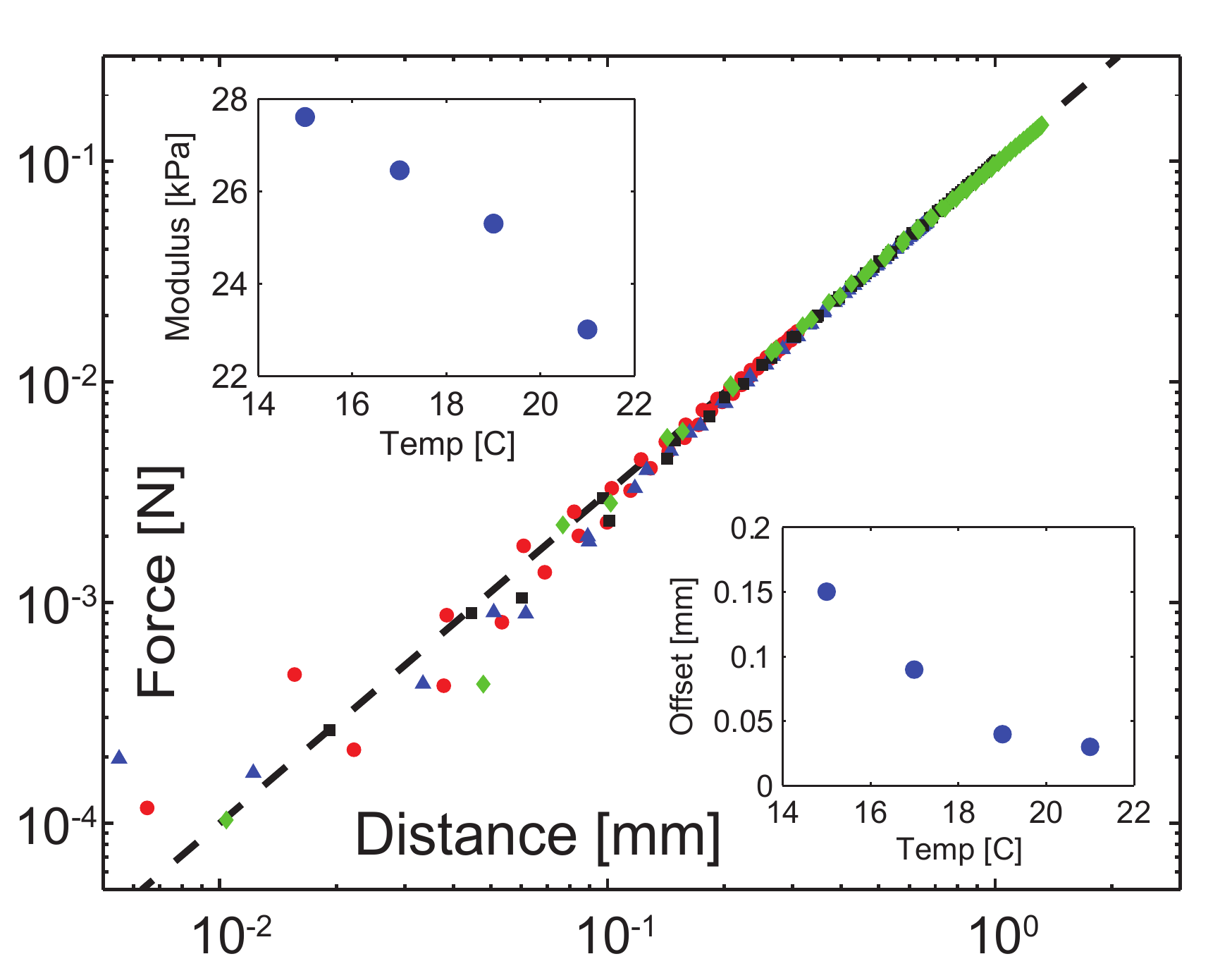}
  \caption{Main panel: force-displacement data from the compression of a submersed hydrogel sphere at $15,17,19,21~^{\circ}C$ (red $\circ$, blue $\triangle$, black $\square$, green $\lozenge$). The dashed line indicates slope 1.5. Vertical and horizontal offsets are removed to achieve a collapse of the data. Top inset: vertical offset from the data in the main panel provides the modulus as a function of temperature. Bottom inset: horizontal offset correction from data in main panel, representing the shrinking of the particle with lower temperature.   \label{fig:tempdepgel}}
\end{figure}

\textit{Temperature dependence ---} The mechanical behavior of hydrogels is generally also temperature dependent. We perform uniaxial compression tests as mentioned above, but then with a $\sim$ 20mm swollen hydrogel particle submersed in a water bath of a specific temperature. Again we measure the expected $F \propto \gamma^{1.5}$ as shown in Fig.~\ref{fig:tempdepgel}. The prefactor in the proportionality was obtained by collapsing the experiments done at different temperatures onto a single master curve. Independent checks~\cite{brodu} had already confirmed that the modulus of the spheres is $23~kPA$ at 21$^{\circ}$ Celsius; this allowed us to determine the modulus of the colder particles by scaling (Fig.~\ref{fig:tempdepgel} inset). The submersed particle also shrinks somewhat during cooling: to quantify this effect, we repeat the same strain cycle at different temperatures and record the offset at which the contact between compressing plate and particle are established. We extract the contact point from the force-displacement data. An increase in the offset of the contact of contact here implies shrinking of the test particle. These offsets are shown in Fig.~\ref{fig:tempdepgel} (bottom right inset). 
 
\subsection{Hydrogel Friction} 
Hydrogels have been shown to have very low friction coefficient with glass or plastic substrates~\cite{slippery1, slippery2, slippery3}. Friction on hydrogels is however rich in phenomenology~\cite{slippery4} including non-Coulomb behavior and rate dependence. We can get an upper bound on the friction coefficient of a hydrogel-hydrogel (HH) contact by stacking two dyed hydrogel cubes (Educational Innovations) on top of each other in a demineralized water bath. The inclination angle at which the cubes start to slide gives us an upper bound for HH friction coefficient of $<0.03$. 
The low friction in HH contacts have a beneficial consequence: since such contacts cannot support shear stresses, a deformation of the contact in a hydrogel sphere is only due to normal forces. Linear elasticity theory then provides a relationship between the contact deformation and the applied normal force. Thus, imaging the deformation of hydrogel spheres allows for the extraction of contact forces in granular packings in three dimensions to a high degree of accuracy.

\subsection{Density \& Index Matching} 
Hydrogel spheres contain 80-99.5\% water by volume, depending on their chemical environment. Thus, the index of refraction of hydrogel spheres is higher, but close to that of water. To image a dense packing of particles, index matching of the fluid phase needs to be fine tuned to that of the spheres to image as many layers as possible~\cite{DijksmanRSI12}. Ordinarily, index matching is achieved by dissolving a second component in the fluid phase. In the case of hydrogel spheres, this approach can fail due to the porosity of the hydrogel. Any small water-soluble molecule, such as glycerol, will diffuse into the hydrogel polymer network, raising both the index of the fluid and the particle. There are two approaches to resolve this. Macromolecules above a particular size are physically too large to diffuse into the polymer network and stay in the fluid phase. We found that polyethyleneglycol (PEG) with a molecular weight of 40,000 diffused into the hydrogel spheres on a timescale of hours, but polyvinylpyrrolidone (PVP) with a molecular weight of 360,000 did not diffuse into the hydrogel over a timescale of weeks. Dynamic light scattering on dilute 360,000 PVP samples indicate that the equilibrium size of these macromolecules is about 20 nanometer.  Note that a concentration gradient of PVP creates an osmotic pressure on the hydrogel spheres, but this is not large enough to result in any visible size change. Adding macromolecules such as PVP to the index matching liquid will however affect the HH friction contact~\cite{ward}; in general the mixing of solvents for index matching can also create non-Newtonian suspensions, as is the case for certain Triton-water mixtures~\cite{beyer}.

In order to image the packing structure, an imaging contrast agent is required. In standard RIMS applications, this is a fluorescent dye that is dissolved in the fluid phase of the system. As mentioned above, the hydrogel spheres are permeable to small molecules, including all fluorescent dyes. A dye dissolved in the fluid phase will therefore also be able to penetrate the hydrogel. Yet, there are some fluorescent dyes that prefer to stay in the hydrogel network rather than in the fluid phase, perhaps due to hydrophobicity~\cite{zanetti}. Gel specific interactions also play a role, as we found that Nile Blue 690 (Exciton) works for commercial hydrogel spheres, but not for custom made hydrogels. For details, see~\citep{dijksman13}.

\section{Imaging and Image Analysis} 

In this section, we discuss some image analysis aspects applicable to soft particles under compression. Soft here means that particles deform \emph{measurably} under mechanical load. This deformation leads to the emergence of asymmetric particles, whose location and contact force extraction require the extraction of their entire surface structure. This is in contrast with most particle tracking codes, which assume a specific (e.g. spherical, elliptical) particle shape. In this part of the article we provide background information on image processing, particle recognition and shape extraction algorithms.

\subsection{Distortion compensation in 3D}
The first step in imaging is to ensure the camera system presents no optical distortion. This is the case with cheap lenses, or by using a zoom lens at suboptimal settings. In order to estimate the distortion field of the camera system, we used the checkerboard method with implementation taken from OpenCV~\cite{opencv,opencv_library}. Images of the checkerboard in multiple 3D orientations are taken in order to estimate both the 3D projection model from the camera to the real world, together with the lens distortion parameters. In practice, for our camera / lens system, no noticeable distortion was observed. 
The second step is to compensate for varying optical path lengths. This is done by measuring the position of each corner of the hydrogel container, in the image for each depth. We thus tag the positions of these corners in selected images at various depths, building a list of $(r,c,1)$ triplets for each corner (standing respectively for row, column, and a constant for homogeneous coordinates). Each of these triplets ideally corresponds to known coordinates $(x,y,z,1)$ in the container referencial. A least-squared error fit is performed~\cite{lourakis04LM} in order to estimate the $3\times 4$ projection matrix $M$ such that
\[
\left\|
\alpha\left[\begin{array}{cccc}
r_{1} & r_{2} &  & r_{n}\\
c_{1} & c_{2} & \ldots & c_{n}\\
1 & 1 &  & 1
\end{array}\right] - M\left[\begin{array}{cccc}
x_{1} & x_{2} &  & x_{n}\\
y_{1} & y_{2} & \ldots & y_{n}\\
z_{1} & z_{2} &  & z_{n}\\
1 & 1 &  & 1
\end{array}\right]
\right\|^2
\] is minimal, where $n$ is the number of observations and  $\alpha$ is a free coefficient for the homogeneous transform.
This method implicitly handles the rotation of the image so that the bottom of the tank appears horizontal. For each image, a robust sub-pixel position of the corners is obtained from the projection. These positions are then used to rescale and stack images into 3D. The net result is a properly scaled three dimensional image based on the (noisy) image slices. The next step is to denoise this 3D image volume.


\subsection{Image Denoising}

\begin{figure}
  \includegraphics[width=1\columnwidth]{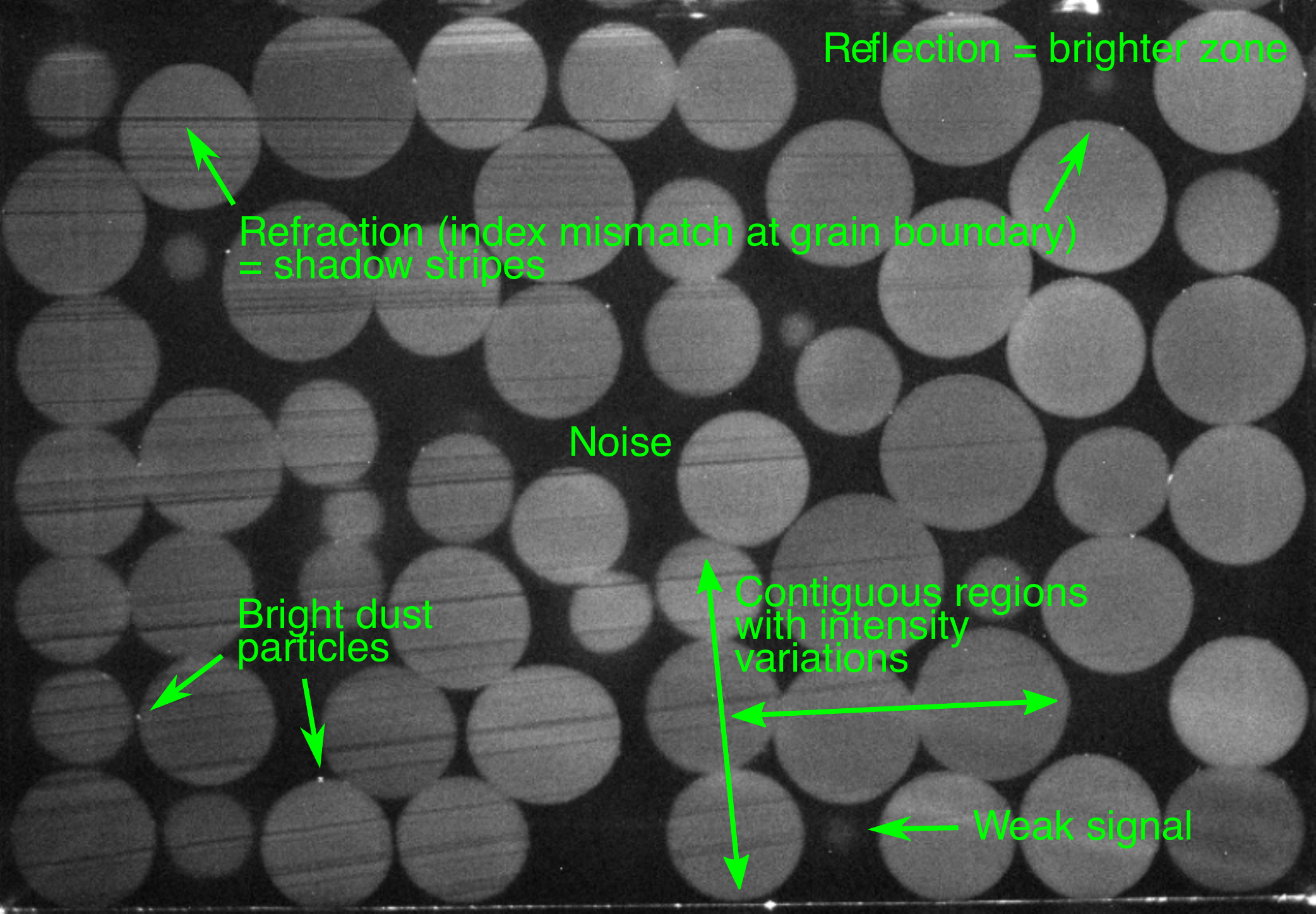}
  \caption{Raw image with major sources of artifacts that needs to be processed.\label{fig:raw_image}}
\end{figure}

\begin{figure}
  \includegraphics[width=1\columnwidth]{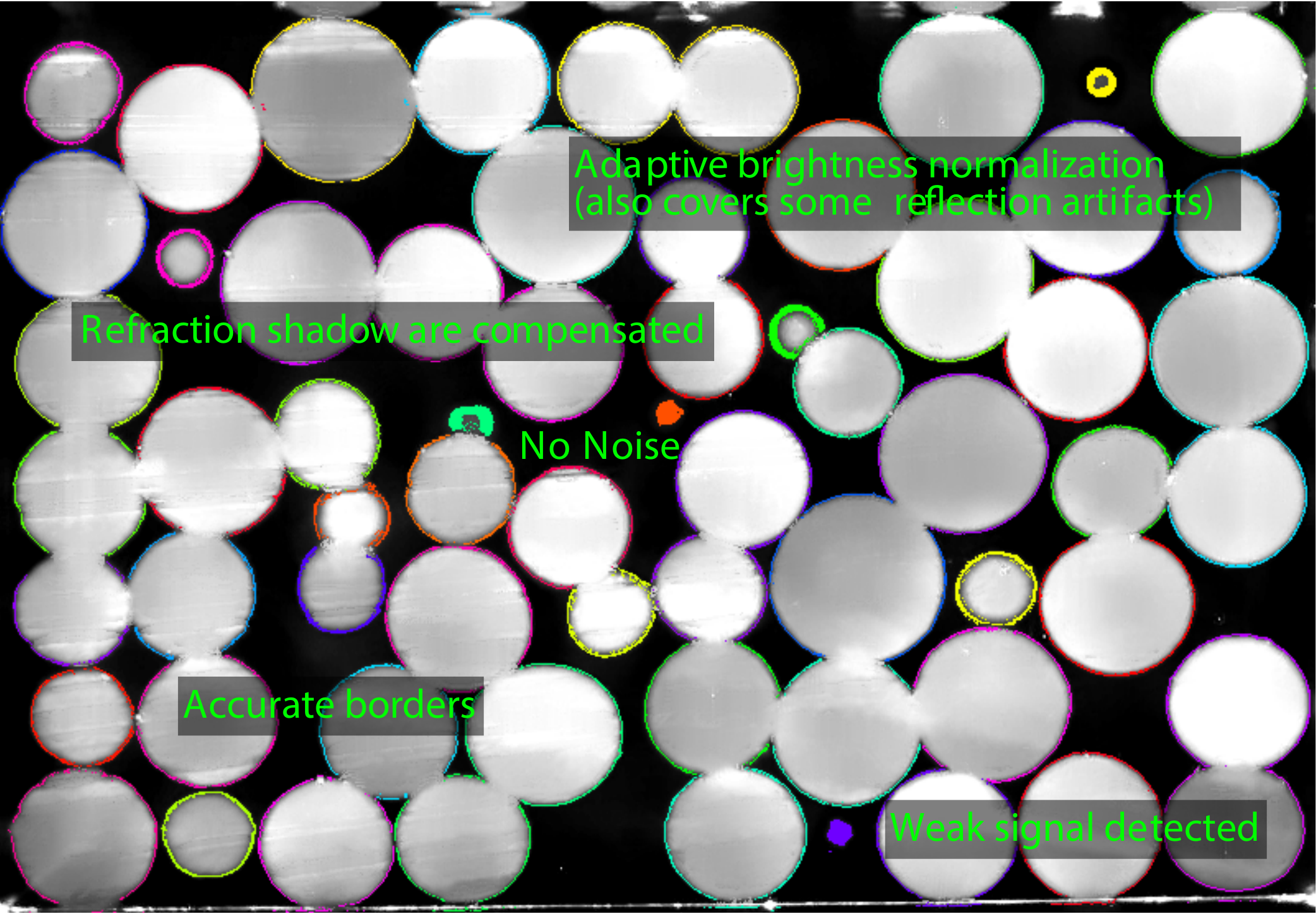}
  \caption{Processed image after artifact removal, and edge detection.\label{fig:processed_image}}
\end{figure}
There are several types of noise in the slice images (shown in Fig.\ref{fig:raw_image}), which we deal with separately (see Fig.\ref{fig:processed_image}). \emph{Dust from particles} and turbidity or leftover dye in water generates noise with unknown characteristics, so we use NL-Means filtering~\cite{nlmeans} to adequately remove it. This step can in fact be performed on the raw images, before stacking them in 3D.\\ 
\emph{Refraction shadows:} These are due to a slight index mismatch, especially at the interface between grains. In the laser sheet plane, the grain cross-section appears as a disk. The deflection on each side of the grain surface produces a dark zone, starting tangentially to the disk, opposite the direction of the laser source. Call D that tangent deflection point. The dark zone starts at D: on the grain inner side the laser is deflected, outside it is not, so the in-between region appears as a thin triangle in the sheet plane. In the plane orthogonal to the laser sheet, but containing at least a laser ray, the geometry is the same and the dark zone has the same shape. In the third orthogonal plane, crossing each laser ray (thus, parallel to the box side opposite the laser source), the dark zones appear as circles. A specific denoising algorithm was created to fill these dark zones. The images are first stacked into 3D to make voxels. That algorithm starts by detecting lines making up the triangles dark zones in one of the first two planes, starting at one grain edge, with a Hough transform~\cite{HoughTransform}. It then fills voxel regions where lines were detected: voxels with the surrounding interstitial fluid ambient intensity are added along the line on the inner side of the dark region. The algorithm then alternates into the other orthogonal plane in order to detect more lines and fill more voxels, until convergence or a prescribed number of iterations were reached. The algorithm could be extended by using circle Hough transforms instead of linear ones and iterating in the third plane, but good results were reached already by alternating in the first two planes.\\ 
\emph{Reflection shadows} and laser light absorption: A local grey-level renormalization is performed. The image is split in overlapping sub-region chunks of $10\times 10$ pixels each. The empirical distribution of gray levels in these chunks is determined, together with its 90\% quantiles interval range. Then, the chunk intensity is renormalized within that range and outside values are clipped to the lower and upper quantile intensity values. This way, differences in intensities on each side of the box due to light absorption are compensated, and difference in illumination due to reflections on elements outside the container are also reduced.

\subsection{Border Voxels}

Finding border voxels at the edge of the grains is a challenge. Starting from the cleaned and preprocessed images, we followed the following steps, detailed hereafter.
\begin{itemize}
\item[-] Finding the grain edge transitions along each 3D axis.
\item[-] Assigning intervals for inner grain voxels between edges.
\item[-] Merging nearby segments to account for false edge detection within grains. 
\end{itemize}

Edge transitions are found by using a smooth numerical differentiator filter~\cite{smoothdiff} on voxel intensities. These filters compute the large variations in intensity while being robust to noise. Fig.\ref{fig:raw_image} shows contiguous regions of intensity, with large variations betwen the grain interiors and exteriors. After this first step, a list of transition voxels is established. Most of these voxels lie on grain surfaces: the dye is concentrated within grains, so interior voxels have higher intensities than the surrounding fluid. Other voxels are outliers, for example due to fine particles in suspension that results in large local intensity variations and thus could not be reduced numerically. 
The next step is to detect these outliers and orient the grain surface voxels by finding grain interiors. For this, we assume that differences in dye absorption are larger between different grains than between different parts of the same grain. With this assumption, a distribution of intensity levels is built along each segment between successive voxels on each 3D axis. Segments from within the same grain should thus sample the same intensity distribution, while segments between different grains should not. A standard Kolmogorov-Smirnov 2-sample test solves this problem, with a confidence threshold empirically determined from experimental data. Similar segments are merged, covering low-intensity regions within grains due to dye inhomogeneity that resulted in spurious transitions. False positives in the merger test may result in intervals larger than a grain diameter and these are discarded. Isolated edge voxels are also removed.

\subsection{Grain attribution and cleaning up voxels}

At this stage, only a small amount of false border outliers remains, small enough to have negligible influence on the grain identification. We then attribute border voxels to unique grain identifiers. In order to do this, a local tangent plane is fit at each voxel, using a Principal Component Analysis on neighbor voxels in order to find the main two directions of largest variation. The plane normal, oriented towards largest voxel intensities, determines the direction toward a rough estimate of the grain center. This proto-center is empirically placed at an average radius distance in this direction. There is one such proto-center per border voxel, and these are considered to be distributed around the true grain center. Assuming these are indeed samples of a probability distribution for the location of the true center, we estimate that distribution with a 3D kernel density estimation \citep{Epanechnikov69}. A simple gradient ascent on that probability distribution is initiated at each proto-center in order to find the local maxima. With a large enough kernel support, the distribution is smooth and only one maxima exists per grain. The location of that maximum is used as an estimate of the grain center.

All border voxels are attributed to the center estimate matching their associated proto-center. Border voxels that are too close or too far from a grain center are considered outliers and eliminated. A voxel reattribution step is then performed in order to deal with limit cases at the interface between two grains. Starting from the estimated grain centers, consider 3D angular sectors spanning all directions. The distribution of border voxels at each distance within the angular sector is estimated by another kernel density estimation. When that density has two peaks, the closest is assumed to point at the true grain surface, and the other to border voxels that were missatributed to this grain. The furthest away voxels are put into a pool of voxels to reassign. Once all grains have been processed, these extra voxels are then reattributed to nearby grain centers, or completely eliminated if that would lead to inconsistent results. This procedure may possibly eliminate a few valid pixels, but it ensures very few outliers or wrong attribution remain.

\subsection{Analytic grain surface estimation}

\begin{figure}
  \includegraphics[width=1\columnwidth]{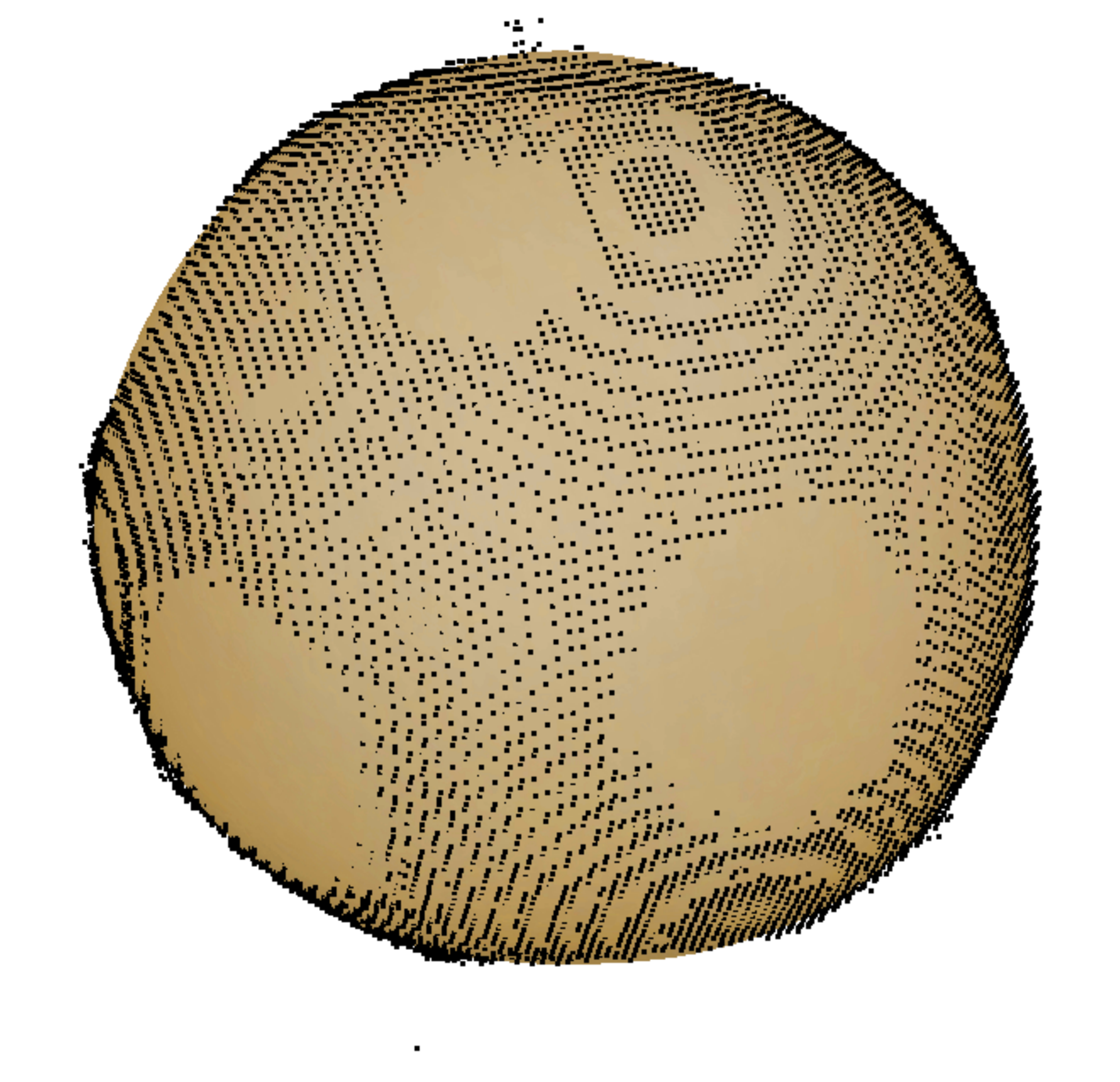}
  \caption{Border voxels with analytic surface fit.\label{fig:border_voxels_and_surface}}
\end{figure}
The goal of this step is to produce an analytic representation of the grain surface, represented as a function that can be evaluated outside border voxels. The process is similar to fitting a curve from measurements, but for a closed surface (shown in Fig.~\ref{fig:border_voxels_and_surface}). Starting from the grain center $G$, a unit vector $\overrightarrow{u}$ defines a direction. The location of the surface in that direction is given by a distance from the grain center $G+f(\overrightarrow{u})\overrightarrow{u}$. The analytic surface representation we choose to use is thus a function $f$  that takes as input a unit vector $\overrightarrow{u}$  in $S^2$, the unit sphere, and produces a positive scalar quantity $f(\overrightarrow{u})\in\mathbb{R}^+$. Triangular spherical B-splines~\cite{strisplines} are used in order to define a basis $B=(b_1,b_2,\ldots,b_n)$ of $n$ functions from $S^2$ to $\mathbb{R}^+$. We then express a grain surface as a set of coefficients in that function basis: $f=c_1b_1+c_2b_2+\ldots+c_nb_n$. The main advantage of this method is that there is no pole, no singularity on the sphere that would result for example from meshing with rectangular longitude/latitude coordinates (or with b-splines in Cartesian coordinates). A second advantage is the isotropy: The triangular spherical B-splines can evaluate any direction in space, but are themselves computed from a set of fixed points in space. These points are taken to be the vertices of a regular geode. The next step is to estimate the surface coefficients $c_{1..n}$ from border voxels  $X_{1..m}$ with $m$ the number of voxels. This is done by least squared error minimization between the border voxels and the estimated surface of the grain: $c_{1\ldots n}=\textrm{argmin}\sum_{i=1}^{m}\left\Vert \left(G+f(\overrightarrow{u})\overrightarrow{u}\right)-X_{i}\right\Vert ^{2}+R$.  A convexity criterion $R$ is used to regularize angular sectors without enough border voxels ( see Fig.~\ref{fig:border_voxels_and_surface}), or to avoid leftover outliers from having too much influence. This criterion stipulates that, for any direction $\overrightarrow{u}$ and directions from neighbor angular sectors $\overrightarrow{v_j}$,  $f(\overrightarrow{u})\overrightarrow{u}$ is within the convex hull of the $f(\overrightarrow{v_j})\overrightarrow{v_j}$ .  This makes the fitting problem non-linear, so we use the Levenberg-Marquardt algorithm~\cite{lourakis04LM}. Once the analytic grain surface is available, we can compute precise geometric parameters from the grain by numerical integration: center of mass, areas of contacts between adjacent grains, etc. In fact, we iteratively reprocess the data using the center of mass instead of the initial center estimate, until convergence for $G$.

\subsection{Computing Forces From Local Surface Structure}

\begin{figure}
  \includegraphics[width=1\columnwidth]{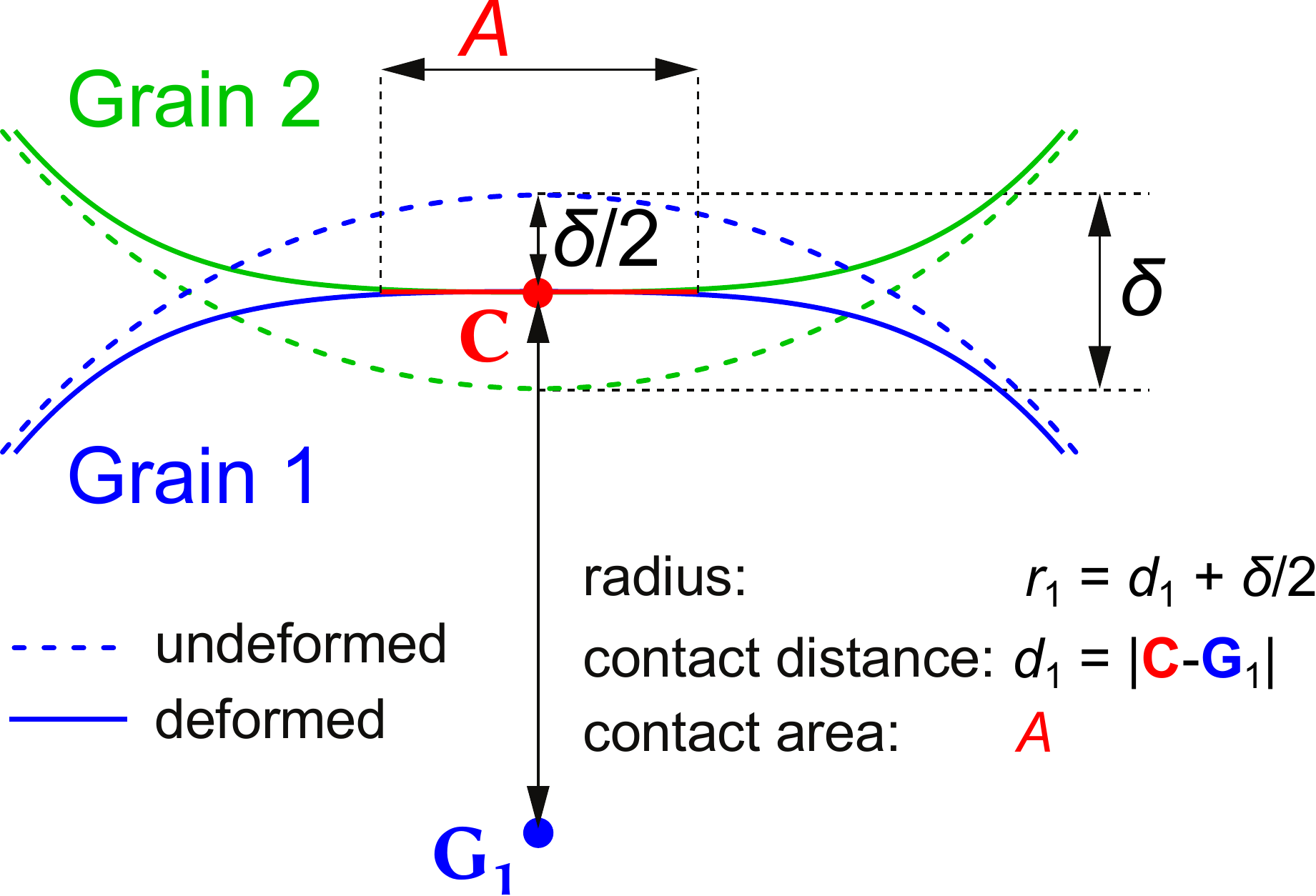}
  \caption{\label{fig:computeforce} Reproduced with permission from Nature Comm. \textbf{6} 6361 (2015) under the Creative Commons Attribution 4.0 International License. Contact and grain geometries, sketched in a 2D plane orthogonal to the contact area. The surface of each grain (blue/green) is shown as solid lines, the hypothetical ``undeformed'' surface of the grains is shown as dashed lines. The deformation, $\delta$, for this contact is the unknown quantity that we infer from linear elasticity. We measure the area of contact $A$, the position of the contact centroid $C$ and its distance $d$ to \textbf{$G_i$}, the center of mass (CoM) of each grain. Note that $d$ (hence $r$) can vary for different contacts on the same particle, as particles are weakly non-spherical. The distance between grain centers $b$ depends mainly on the grain polydispersity and weakly on their non-sphericity so, on average, $\langle b \rangle \sim 2\langle d \rangle$. Our force inference algorithm uses the observed surfaces, without assuming sphericity, and provides experimental access to all the relevant quantities in order to reconstruct $\delta$.}
\end{figure}

We use the general theory of linear elasticity~\cite{LnL} in order to derive force values from grain deformations. Observed images correspond to deformed grains, so the first step is to derive the deformation $\delta$ from the undeformed surface. With all grains having the same Young's modulus, $\delta$ is split equally at each contact (Fig.~\ref{fig:computeforce}). Thanks to the analytic surface representation from the previous section, we know the position of the center of mass $G$ and of the center of the contact area $C$. Therefore, the undeformed radius of curvature is $r = d + \frac{1}{2}\delta$, with $d$ the distance between $G$ and $C$. The force is given both by $F = E_e r^{1/2}\delta^{3/2}$ and by $F = E_e\delta a$, with $E_e$ the effective Young's modulus, $\frac{1}{r_e} = \frac{1}{r_1} + \frac{1}{r_2}$ the effective contact radius and $a = \sqrt{A/\pi}$ the radius of the area of contact. The effective Young's modulus appears in both expressions and does not intervene for deriving $\delta$ from purely geometric measurements: $r^{1/2}\delta^{3/2}=\delta a$ is obtained by equating both force expressions. This leads to a cubic equation for $\delta$, which we solve at each contact. Once $\delta$ is found, we use an independent measurement of the Young's modulus, $E \sim 23~kPa$ for our particles, in order to convert $\delta$ into a force value. Assuming negligible friction, the 3D force vector points in the direction $G-C$ and is thus available at each contact.

\subsection{Coarse Graining the Stress Tensor}

The goal of this section is to validate our method and embedded assumptions, for example the use of Hertz' law for the contact force, by comparing the image-reconstructed global force value acting on a compressed and imaged packing with the force on the top plate that is compressing the imaged packing. This independent force measurement represents a third independent measurement, the first two being the 3D images themselves and the Young's modulus measurement on a single compressed sphere. Unfortunately, directly summing the force values of individual contacts on the top plate is not possible due to image processing inaccuracies at grain/plate boundaries. Finding the precise location of that top plate is difficult for the same reason.

The solution is to coarse-grain individual contacts between grains in order to get a continuous expression of the stress tensor, which is then integrated over the top plate in order to get the global force value to compare with the measurement. The stress tensor can indeed  be integrated over a plane at any height $h$  within the tank, leading to a force $F(h)$. We use the hydrostatic pressure gradient in order to find the precise location of the top plate within the tank referential, by selecting the $h$  for which $F(h)$  equals the measured force.

We compute the coarse-grained continuum stress tensor from the average balance equations of~\cite{Babic97}. Assuming no dynamic components, these simplify and can be computed purely from geometric properties of the grains, which are available as detailed in the previous sections. Spatially continuous values, outside contact points, are derived by applying a kernel density estimation~\cite{Epanechnikov69}. The coarse-graining scale, hence kernel bandwidth, is set to be one average grain diameter. We thus compute continuous expressions of the stress tensor at the scale of grains. This coarse graining correctly computes the force dynamics on a compressed particle packing; for the numerical evidence we refer the reader to Ref.~\cite{brodu}.

\subsection{Accuracy of Image Analysis}

\begin{figure}
  \includegraphics[width=1\columnwidth]{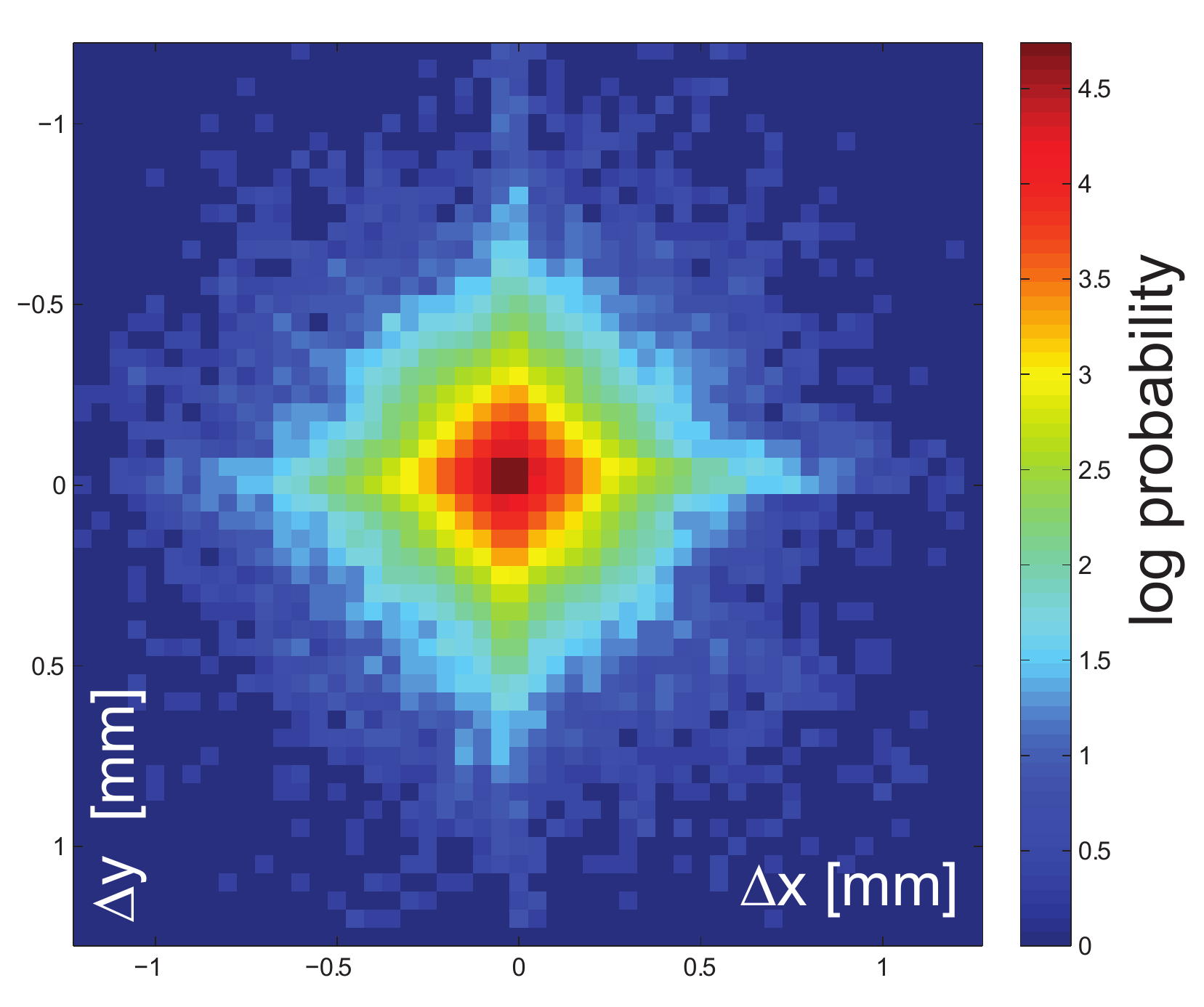}
  \caption{Probability of finding a CoM displacement in the $\textbf{x},\textbf{y}$-plane during $\textbf{z}$ compression. \label{fig:xydisp}}
\end{figure}

\textit{Displacement resolution} --- In order to quantify the accuracy of the image analysis methods we described above, we look at two different metrics. First, we probe the resolution of the center of mass (CoM) detection through measuring displacement fields in a uniaxially compressed packing of hydrogel spheres. If the compression direction is in the $\textbf{z}$ direction, the displacements of the CoMs of the particles in the packing in the $\textbf{x},\textbf{y}$ directions should be minimal and uncorrelated. We plot the distribution of these displacements in Fig.~\ref{fig:xydisp}. We can clearly see that the observed displacements in the $\textbf{x},\textbf{y}$ directions per $\textbf{z}$-compression step are mostly very small. The standard deviation of the distribution of both $\textbf{x}$ and $\textbf{y}$ displacements is within 50 micron. Moreover, the distribution has azimuthal symmetry, with only minor preferences for detecting displacements along the coordinate axes.\\

\begin{figure}
  \includegraphics[width=1\columnwidth]{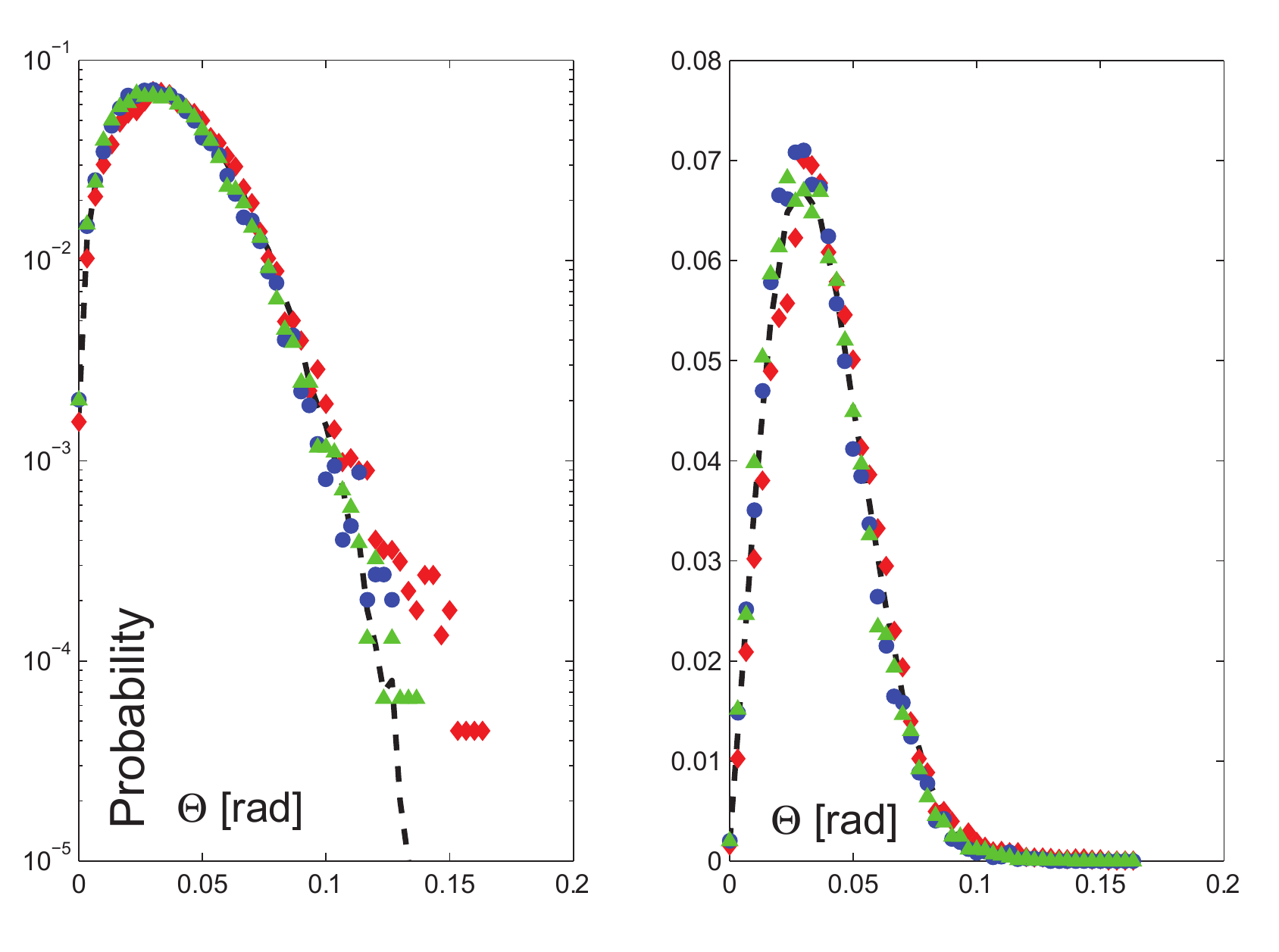}
  \caption{(a) histogram of angles between CoM and CoB for a strain of $\sim 15\%$ (red $\lozenge$) and $\sim 3\%$: blue $\circ$ (during compression); green $\triangle$ (during decompression). The dashed line indicates a distribution of angles obtained with a location detection noise of 30~microns. (b) Same data on linear scale.}\label{fig:anglenoise}
\end{figure}

\textit{Angular resolution} --- The calculation of stress tensors from the microscopic information in the images involves a large number of dot products and other vectorial analyses. It is thus crucial to get a sense of the accuracy of angle measurements in the packing. To estimate the angular resolution of the microscopic data, we quantify the angle between the vector connecting the CoMs of two contacting particles and the branch vector to their contact,  so between \textbf{$G_i$}-\textbf{$G_j$} and \textbf{$G_i$}-\textbf{$C$} as indicated in Fig.~\ref{fig:computeforce}. This angle $\theta$ should in principle be zero for perfectly spherical particles, but will deviate due to inaccuracy in CoM determination and, at higher compression levels, perhaps due to pronounced asymmetry of particle shape. The angle inaccuracy due to noise in CoM location follows a modified Rayleigh distribution as follows: noise in an angle between vector $i$ and $j$ whose expected deviation is null can be quantified by expressing $j$ as $\epsilon_x,\epsilon_y$ components in the plane of which $i$ is the normal hence aligned with the \textbf{z}-axis. This leads to angle noise of $\cos(\theta) = 1/\sqrt{\epsilon_x^2 + \epsilon_y^2 + 1}$, assuming that $i$ is normalized, and the $j = \{\epsilon_x, \epsilon_y, 1\}$. For $\epsilon_{x,y}$ Gaussian variables, this produces a modified Rayleigh distribution. In Fig.~\ref{fig:anglenoise} we plot this distribution for a set of uncorrelated Gaussian distributed $\epsilon_{x,y}$ with a standard deviation of 30~micron (dashed line). This CoM location detection noise is consistent with the displacement noise shown in the same Fig.~\ref{fig:xydisp} and captures the experimental data on angle noise very well. Only at maximum compression can a small deviation be observed, perhaps due to pronounced asphericity of the particles at this compression levels. We conclude that vectorial angle noise is due to CoM and contact location detection noise.

\section{Conclusions}
We discussed the use of hydrogel spheres for their use in index matched scanning experiments on the mechanics of three dimensional granular samples. We discussed both index matching aspects and image analysis techniques that allow one to extract the aspherical surface structure of compressed particles inside a mechanically deformed packing. The surface structure data so extracted gives access to the particle stress tensor and the moment of inertia of a particles, which are relevant microscopic ingredients for any micromechanically tested theory of granular media. The source code for the methodology described in this paper, together with the data, are available at http://nicolas.brodu.net/recherche/tomorim/. We thank Matthias Schr\"{o}ter and Thorsten P\"{o}schel for their hospitality during the Erlangen Spring School in April 2016. 

\bibliographystyle{aipproc}   



\begin{thebibliography}{42}
\bibitem{footnote1}
 These particles are sold under names such as Soil Moist, Aquagel,
  Aquabeads, Decobeads, et cetera. The composition is a proprietary formula
  from JRM Chemicals.
\end{thebibliography}


\end{document}